\documentclass[12pt,preprint]{aastex}
\usepackage{graphicx}

\newcommand{\asca}{{\small \it ASCA}}

\newcommand{\rosat}{{\small \it ROSAT}}

\newcommand{\agn}{{\small AGN}}
\newcommand{\ngc}{{\small NGC~3783}}
\newcommand{\iras}{{\small IRAS~13349+2438}}
\newcommand{\uta}{{\small UTA}}
\newcommand{\hullac}{{\small HULLAC}}
\newcommand{\xstar}{{\small XSTAR}}
\newcommand{\fwhm}{{\small FWHM}}

\newcommand{\xmm}{{\it XMM-Newton}}
\newcommand{\rgs}{{\small RGS}}
\newcommand{\chandra}{{\it Chandra}}

\newcommand{\hetgs}{{\small HETGS}}

\newcommand{\uv}{{\small UV}}
\newcommand{\x}{X-ray}

\newcommand{\cmsq}{cm$^{-2}$}
\newcommand{\kms}{km~s$^{-1}$}

\begin{document}

\title{Absorption Measure Distribution of the Outflow in IRAS~13349+2438:
Direct Observation of Thermal Instability?}
\author{Tomer Holczer \altaffilmark{1}, Ehud Behar \altaffilmark{1} and
Shai Kaspi \altaffilmark{1},\altaffilmark{2}}

\altaffiltext{1}{Department of Physics,
                 Technion, Haifa 32000, Israel.
                 tomer@physics.technion.ac.il (TH),
                 behar@physics.technion.ac.il (EB),
                 shai@physics.technion.ac.il (SK).}
\altaffiltext{2}{School of Physics and Astronomy, Raymond and
Beverley Sackler Faculty of Exact Sciences, Tel Aviv University,
Tel-Aviv, 69978 ISRAEL }

\received{} \revised{7/1/04} \accepted{}

\shorttitle{AMD of \iras} \shortauthors{Holczer et al.}

\begin{abstract}
We analyze the \chandra\ \x\ spectrum obtained with the \hetgs\
grating spectrometer of \iras, which has one of the richest
absorption spectra of a quasar outflow. Absorption from almost all
charge states of Fe is detected. This allows for a detailed
reconstruction of the absorption measure distribution ($AMD$),
which we define as the continuous distribution of column density
as a function of ionization parameter. We find a double peaked
$AMD$ for \iras\ with a total (ionized) column density of $N_H$ =
(1.2$\pm$0.3)$\times10^{22}$~\cmsq\ assuming solar iron abundance.
For comparison, we perform a similar analysis on the well studied
\hetgs\ spectrum of \ngc.
% for which we find $N_H$ =(4.1$\pm$0.7)$\times10^{22}$~\cmsq.
Both sources feature a deep minimum in column density consistent
with no absorption from gas at temperatures of $4.5 < \log T < 5$
(K). We interpret the minima as observational evidence for thermal
instability in this temperature regime.

\end{abstract}


\section{INTRODUCTION}
\label{sec:intro}

The \x\ spectra of many active galactic nuclei (\agn s) viewed
directly toward the central source (e.g., Seyfert 1 galaxies) show
the continuum flux absorbed by numerous absorption lines produced
by ionized gas. The lines are generally shifted to shorter
wavelengths, indicating an outflowing wind. It is not clear
whether this wind can also be associated with the line-of-sight
material responsible for the optical reddening of these sources.
To that end, the outflow would have to entail neutral- gas or dust
grains along with the highly ionized wind. It is still
controversial whether the winds are significant for the \agn\
central engine or for the host galaxy in terms of mass, energy,
and momentum. It is possible that extreme outflows may even play a
central role in cosmological feedback and in the metal enrichment
of the intergalactic medium (IGM). In order to address these open
questions, it is important to develop reliable techniques to probe
the winds and to perform accurate measurements of their physical
properties.

The \x\ band is advantageous for \agn~outflow studies not only
because most of the outflow appears to be highly ionized, but also
since the \x\ band comprises detectable absorption lines from the
full range of charge states: From neutral up to hydrogen-like.
Moreover, in most cases an \x\ spectrum features several lines
from a single ion, which helps to deduce ionic column densities
from saturated absorption lines. The detection of many charge
states of a single element is key to plasma diagnostics as it
allows to model the distribution of ionization and to measure the
total column density independent of assumptions regarding the
relative element abundances.  For this purpose, Fe ions are
particularly useful as they form over several orders of magnitude
of the ionization parameter, defined \citep[e.g.,][]{kal95} by

\begin{equation}
\xi=\frac{L}{n_Hr^2}
\end{equation}

\noindent where $L$ is the ionizing luminosity between 1 and 1000
Rydberg, $n_H$ denotes the hydrogen number density and $r$ is the
distance of the absorber from the ionizing source.  Once the
distribution of ionization is known for one element, the relative
abundances of other elements can be measured by comparison. For
the absolute abundances of metals with respect to hydrogen, one
still needs to utilize \uv\ spectra, as the energies of atomic
transitions in hydrogen are too low to be observed in the \x s.
Highly ionized K- and L- shell ions (1-2, and 3-10 electrons,
respectively) of Fe appear regularly in astrophysical \x\ sources,
either in emission or in absorption. The lower charge states of Fe
(11-26 electrons) feature strong inner-shell absorption lines in
the soft X-ray band \citep{uta01}. Each ion can produce dozens of
overlapping lines forming unresolved transition arrays (\uta s).
Interestingly, \iras\ was the first source in which Fe M-shell
\uta s were observed \citep{sako01}.

The majority of works on the \x\ spectra of \agn\ outflows fit the
data with a gradually increasing number of iso-$\xi$ absorption
components, until the fit is satisfactory \citep[e.g.,][]{kaspi01,
kaastra02,netzer03}. For high-quality spectra two or three
ionization components produce a good fit. In this work, we employ
a more fundamental method, which tries to reconstruct the actual
distribution of the column density in the plasma as a continuous
function of $\xi$. We call this distribution the absorption
measure distribution ($AMD$). The present method is similar in
principle to that tested by \citet{katrien05} and by
\citet{elisa06}, who assumed a power-law distribution. However,
here we fit for the $AMD$ directly from the measured ionic column
densities making no {\it a-priori} assumptions regarding its
shape.

\iras\ is a bright type-1 quasar ($L~> 10^{46}$ erg~s$^{-1}$) at a
redshift of z=0.10764 \citep{kim95}. Optical and infrared
%spectral and polarization
observations by \citet{wills92} have suggested absorption of the
central source by dust. \x\ observations with \rosat\
\citep{brandt96} and \asca\ \citep{brandt97} did not find a
significant column density of cold material, but suggested the
presence of ionized gas along the line of sight, perhaps with
embedded dust grains. \citet{siebert99} restated the difficulty to
explain the optical and \x\ observations with the same absorber.
The line-resolved \rgs\ (Reflection Grating Spectrometer on board
\xmm) spectrum of \iras\ measured by \citet{sako01} revealed broad
(\fwhm\ $\simeq\ 1400 \pm$ 300 \kms) absorption lines of many
ions. Two distinct ionization components were determined at $\log
\xi \simeq 0$ and at $\log \xi \simeq 2 - 2.5$ where $\xi$ is in
c.g.s. units (erg cm s$^{-1}$). The low- and high- $\xi$
components were found to have column densities of a few 10$^{21}$
\cmsq\ and a few 10$^{22}$ \cmsq, respectively. The good agreement
between the low-$\xi$ column density and that deduced from the
optical reddening \citep{wills92} led \citet{sako01} to conjecture
that the low-$\xi$ absorber might be responsible also for the
optical reddening. In this paper, we revisit the ionized absorber
of \iras\ using a long exposure with the High Energy Transmission
Grating (\hetgs) spectrometer on board \chandra\ performed 3 years
and 9 months after the \xmm\ observation. The spectral resolution
of \hetgs\ is much better than that of the \rgs\ and the long
exposure compensates for its inferior effective area.

\section{DATA REDUCTION}
\label{sec:data}

\iras\ was observed by \chandra /\hetgs\ on 22--26 February, 2004
for a total exposure time of 300~ks. All observations were reduced
from the \chandra\ archive using the standard pipeline software
(CIAO version 3.2.1). The total number of counts in the first
(plus and minus) orders between 2 and 20~\AA\ is 22836 for MEG
(medium energy grating) and 9542 for HEG (high energy). No
background subtraction was required as the backdround level was
negligible. The post-pipeline procedure is similar to the method
reported in \citet{kaspi01}. Flux spectra were obtained by
dividing the count spectra by the effective area curves produced
by the standard CIAO tool. Positive and negative first-order data
of both the MEG and HEG instruments were combined.  The spectrum
was corrected for neutral galactic absorption \citep[$N_H$ =
1.1$\times10^{20}$ \cmsq,][]{murphy96}. The data were rebinned to
36~m\AA\ bins to improve the statistics of the spectra with
minimal compromise of the spectral resolution, as the MEG
instrumental line spread function is approximately 23~m\AA\ broad
(\fwhm).

\section{SPECTRAL MODEL}
\label{sec:modeling}

Variations of approximately 25\% on time scales of 20~ks were
observed during the 300~ks exposure of \iras. Due to this low
level of variability and since we are interested in the over all
properties of the ionized absorber, henceforth, we use only the
combined MEG and HEG full 300~ks spectrum. The present fitting
procedure follows our ion-by-ion fitting method \citep{behar01,
behar03, tomer05}. First, we fit for the broad-band continuum.
Subsequently, we fit the absorption features using template ionic
spectra that include all of the absorption lines and photoelectric
edges of each ion, but vary with the broadening (so-called
turbulent) velocity and the ionic column density.

\subsection{Continuum Parameters}
\label{sec:continuum}
\begin{figure}
%  \plotone{Varibility.eps}
% \centerline{\psfig{figure=NeIX.eps,height=7in}}

\centerline{\includegraphics[width=13cm,angle=90]{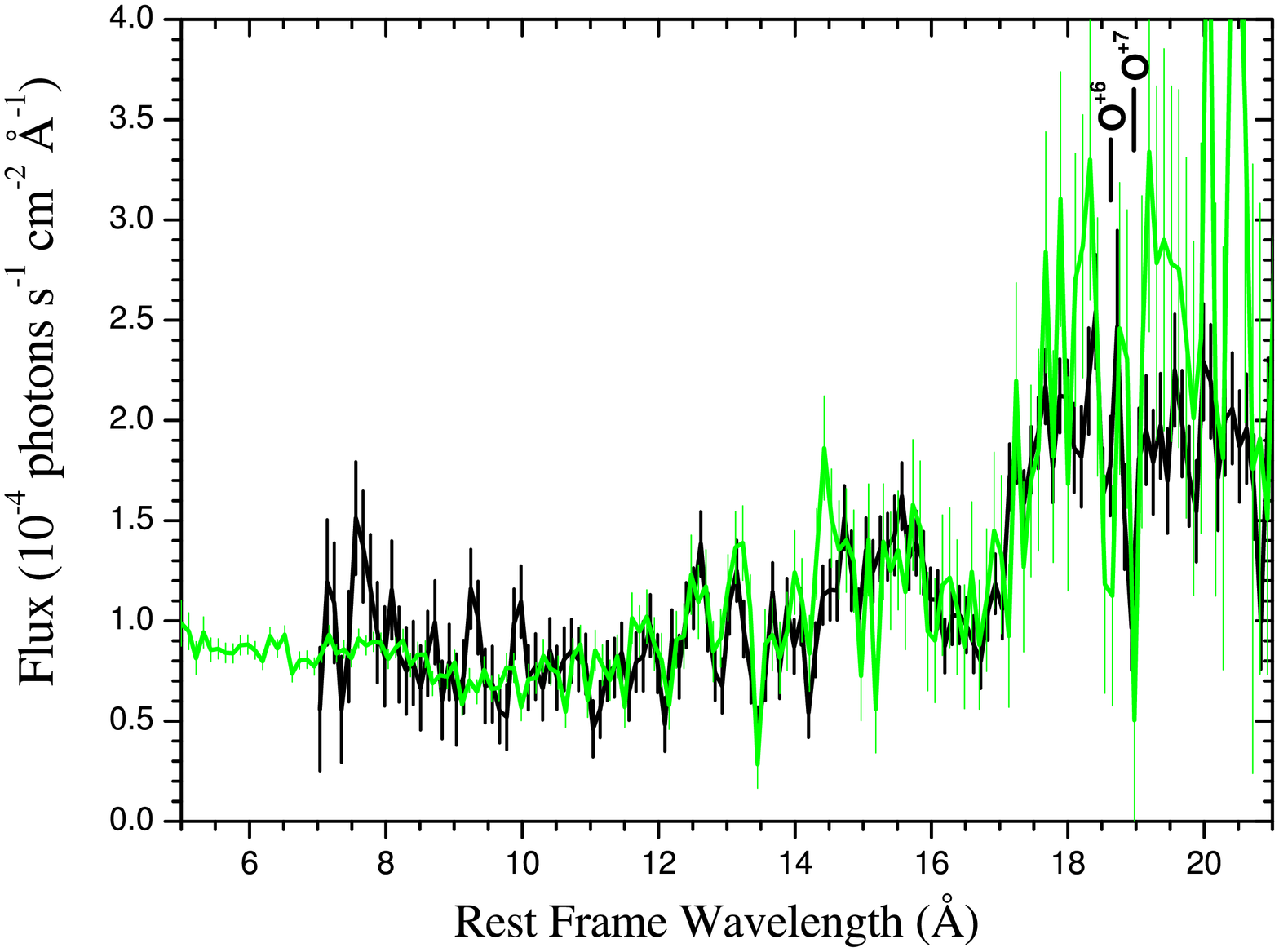}}

  \caption{The present \hetgs\ spectrum of \iras\ (in green) observed on 22--26 February, 2004 compared to the \rgs\
  spectrum (in black) observed on 19--20 June, 2000.
  Despite the $\sim$~4 years difference, the flux level of the two observations appears
  almost unchanged, especially the hard X-Ray region and up to 17~\AA.
  There is noticeable variability beyond 17~\AA. Two O line locations are marked to demonstrate
   the absorption variability of oxygen.
   }
   \label{plotone}
\end{figure}

 The continuum spectrum of \iras\ can be
characterized by a high-energy power-law and a blackbody that
rises at lower energies. The present best-fit power-law parameters
are a photon index of $\Gamma_\epsilon$=~1.9 with a total flux in
the 5--35 \AA\ range of 2.8$\times10^{-3}$ photons cm$^{-2}$
s$^{-1}$, to be compared to $\Gamma_\epsilon$= 2.2 and a flux of
2.4$\times10^{-3}$ photons cm$^{-2}$ s$^{-1}$ obtained by
\citet{sako01}.
% Note that the wavelength power-law photon index $\Gamma_\lambda$
%distribution $F_\lambda~=~A(\frac{\lambda}{12.4})^{-\Gamma_\lambda}$. The
%wavelength powerlaw index $\Gamma_\lambda$ that
%is related to the energy power-law photon index $\Gamma_\epsilon$
%through $\Gamma_\lambda=~2-\Gamma_\varepsilon$.
The current
blackbody has a temperature of $kT=$ 105~eV and a 5--35 \AA\ flux
of 7.8$\times10^{-3}$ photons cm$^{-2}$ s$^{-1}$, compared to
$kT=$ 100~eV and 4.0$\times10^{-3}$ photons cm$^{-2}$ s$^{-1}$
obtained by \citet{sako01}. The continuum applied by
\citet{brandt97} to the \asca\ data produced a flux in the 2--10
keV range of 5.7$\times10^{-12}$ erg cm$^{-2}$ s$^{-1}$ compared
to the present value of 3.6$\times10^{-12}$ erg cm$^{-2}$
s$^{-1}$.
 The continuum applied by \citet{brandt96} to the \rosat\ data produced
 a flux in the 0.1-2.5 keV range of 1.2$\times10^{-11}$ erg cm$^{-2}$ s$^{-1}$
compared to the present value of 1.7$\times10^{-11}$ erg cm$^{-2}$
s$^{-1}$. In short, the continuum measured here with \chandra\ is
within $\sim$50\% of all previous observations. In particular, the
continuum model parameters are fairly similar to those of
\citet{sako01}.
% implying that the central
%source of \iras\ did not drastically vary over the four years that
%separate the two observations.
In Fig.~\ref{plotone} we present a direct comparison of the two
grating spectra from these two observations showing their over all
similarity. It can be seen that the hard \x\ flux is almost
identical, while there is some increase in the soft flux. This
change is reflected in the small difference between the two models
in the blackbody normalization parameters. One might notice that
the depth of the O K-shell edges is also different between
observations. This is discussed in more detail below in \S
\ref{sec:results}.

\subsection{The Ionized Absorber}

The spectrum $I_{ij}(\nu )$ around an atomic absorption line $i
\rightarrow j$ can be represented by:

\begin{equation}
I_{ij}(\nu) = I_0(\nu)~{\mathrm exp}~[-N_{ion}\sigma_{ij}(\nu)]
\end{equation}

\noindent where $I_0(\nu)$ represents the continuum intensity,
$\sigma_{ij}(\nu)$ denotes the line absorption cross section for
photoexcitation (in cm$^2$) from ground level $i$ to level $j$. If
all ions are essentially in the ground level. $N_{ion}$ is the
total ionic column density towards the source in cm$^{-2}$. The
photoexcitation cross section is given by:

\begin{equation}
\sigma_{ij}(\nu) = \frac{\pi e^2}{m_ec}f_{ij}\phi(\nu)
\end{equation}

\noindent $f_{ij}$ denotes the absorption oscillator strength and
$\phi(\nu)$ represents the Voigt profile due to the convolution of
natural (Lorentzian) and Doppler (Gaussian) line broadening. The
Doppler broadening consists of thermal and turbulent motion, but
in \agn~outflows, the turbulent broadening appears to totally
dominate the temperature broadening. Transition wavelengths,
natural widths and oscillator strengths were calculated using the
Hebrew University Lawrence Livermore Atomic Code
\citep[\hullac,][]{bs01}. More recent and improved atomic data for
the Fe M-shell ions were incorporated from \citet{gu06}.

\begin{figure}
%  \plotone{fe16.eps}
% \centerline{\psfig{figure=NeIX.eps,height=7in}}

\centerline{\includegraphics[width=13cm,angle=90]{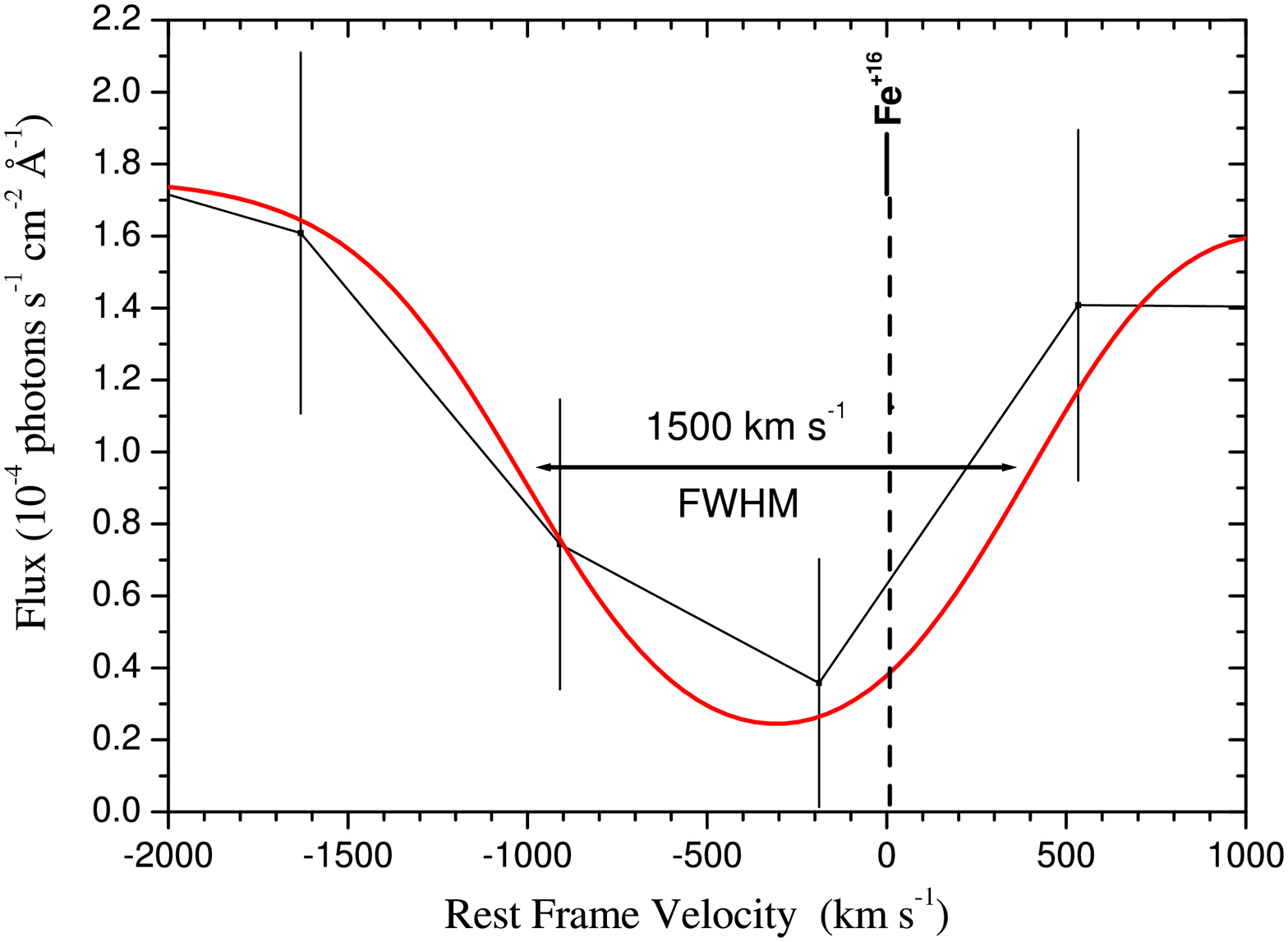}}

  \caption{
  Profile of the strong, isolated Fe$^{+16}$ absorption line at 15.01~\AA\
  plotted in velocity space with the best-fit profile in red.
  The global best-fit position of $-300$~\kms\ (blue-shift)
  and best-fit broadening corresponding to \fwhm\ of 1500~\kms\
  ($v_{turb} = 640$~\kms) are demonstrated. }
   \label{plottwo}
\end{figure}

The Doppler broadening can be typically characterized by a
turbulent velocity $v_{turb}$, although for very high resolution
\uv\ spectra of \agn\ outflows, this approximation has been shown
to be inadequate and velocity-dependent partial covering effects
need to be taken into account \citep{arav99}. Due to the limited
resolution, the present \hetgs\ spectrum of \iras\ does not
warrant a more sophisticated analysis than the $v_{turb}$
approximation. Indeed, a turbulent velocity of $v_{turb}$~= 640
\kms, corresponding to a \fwhm\ of $\sim$ 1500 \kms, provides a
good fit to the strongest absorption lines in the spectrum as
demonstrated in Fig.~\ref{plottwo} for the Fe$^{+16}$ absorption
line at 15.01~\AA. This specific absorption line is shown because
it has a high oscillator strength of 2.3 and it is relatively
unblended. The value of $v_{turb}$~= 640 \kms\ is consistent with
that quoted by \citet{sako01}.

Since the absorbing gas is outflowing, the absorption lines are
slightly blue-shifted with respect to the \agn\ rest frame. The
current best-fit outflow velocity is $-300~\pm ~50~$\kms. Errors
correspond to $\Delta \chi^2 = 1$. \citet{sako01}, on the other
hand, found from the \rgs\ spectrum two velocity components, high-
and low- ionization components with velocities of
+20$~^{+330}_{-200}$~ \kms, and --420$~^{+180}_{-190}$~ \kms,
respectively. We find that a single outflow velocity of $-300~\pm
~50$~\kms\ actually provides a better fit to the \hetgs\ spectrum.
In Fig.~\ref{plottwo}, the centroid of the Fe$^{+16}$ line can be
seen to be offset from the zero-velocity position by at least six
sigma ($\sigma = 50$~\kms). \citet{sako01} ascribed this line to
the high-ionization component with $v_{out}$=
+20$~^{+330}_{-200}$~ \kms. The $-$300 \kms\ blue-shift (obtained
by fitting {\it all} absorption lines in the spectrum
simultaneously) clearly provides a better fit to the profile of
this specific line. We ascribe the potential discrepancy with
\citet{sako01} to the superior spectral resolution of the present
\hetgs\ measurement and to the large errors on the velocities of
\citet{sako01}. If anything, the data in Fig.~\ref{plottwo} hint
at an even higher velocity for this specific line, which might
suggest that there is some scatter in velocity between the
different ions. However, this effect is too weak for us to measure
even with \hetgs\ and henceforth, we assume a uniform outflow
velocity of $-300~\pm ~50$~\kms.

\begin{figure}
  \centering
  \includegraphics[width=13.0cm]{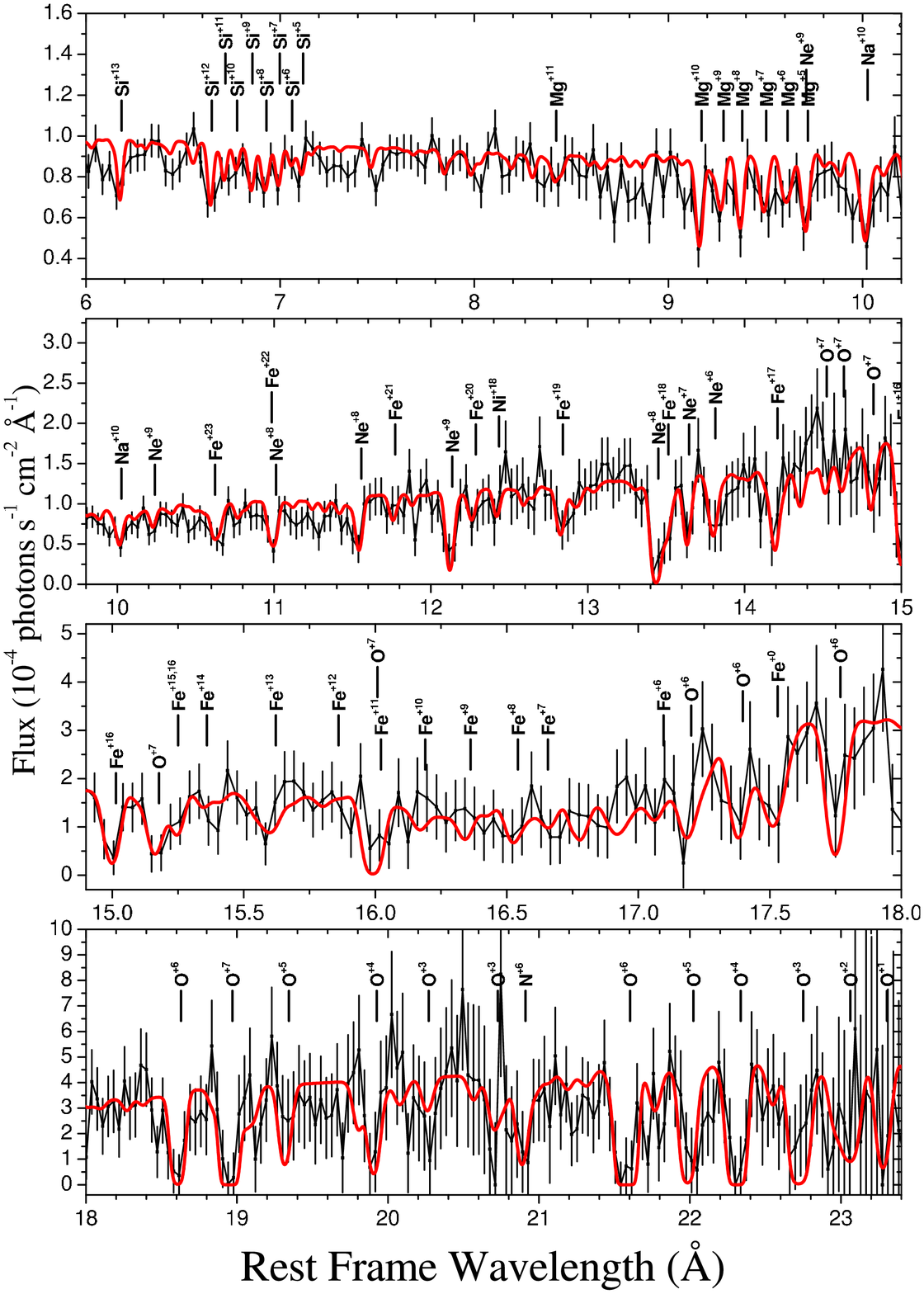}

  \caption{
  \textit{Chandra} \hetgs\ spectrum of \iras\ corrected for cosmological redshift ($z$ = 0.10764).
   The red line is the best-fit model in which a blue-shift velocity of --300
   \kms\ was applied to all ions. Ions responsible for the
   strongest lines and blends are marked above the data.
%    and a turbulent velocity v$_{turb}$~=~640 \kms~ was used.
}
  \label{plotthree}
\end{figure}

Our model includes all of the important lines of all ion species
that can absorb in the waveband observed by \hetgs. In the present
spectrum of \iras, we find evidence for the following ions:
N$^{+6}$, all charge states of O, Ne$^{+6}$--~Ne$^{+9}$,
Mg$^{+4}$--~Mg$^{+11}$, Na$^{+10}$, Si$^{+5}$--~Si$^{+13}$, 21
ions of Fe, and Ni$^{+18}$. We also include the K-shell
photoelectric edges for all these ions although their effect here
is largely negligible. When fitting the data, each ionic column
density is treated as a free parameter. A preliminary fit is
obtained using a custom made Monte-Carlo fitting method applied to
the entire spectrum. Subsequently, the final fit is obtained for
individual ionic column densities in a more controlled manner,
which ensures that the fit of the leading lines is not
compromised. The best fit model for the present spectrum of \iras\
is shown in Fig.~\ref{plotthree}. It can be seen that the Si, Mg,
Ne, and Fe L-shell lines are reproduced fairly well by the model.
However, in the Fe M-shell region between 15.0 -- 17.5 \AA, the
spectrum is rather noisy and it is difficult to identify
individual ions. For these ions, only rough estimates of the
column densities can be obtained. Note that some lines could be
saturated, e.g., the leading lines of O$^{+6}$ and O$^{+7}$. In
these cases, the higher order lines with lower oscillator
strengths are crucial for obtaining reliable $N_{ion}$ values.

\section{IONIC COLUMN DENSITIES}
\subsection{Results}\label{sec:results}

The best-fit ionic column densities are listed in
Table~\ref{tab1}. For the most part, the column densities of the
Fe, Si, and Mg ions are of the order of
10$^{16}$--10$^{17}$~cm$^{-2}$. The column densities of N$^{+6}$
and the Ne ions exceed 10$^{17}$~cm$^{-2}$, while those of the O
ions reach $\sim$~10$^{18}$~cm$^{-2}$. Comparing these results
with those of \citet{sako01}, we find that the Fe M-shell column
densities are consistent. The present column densities for the O
K-shell ions are much higher, while those of Fe L-shell are much
lower than those of \citet{sako01}. In order to understand this
discrepancy, we went back to the \rgs\ spectrum and compared both
data and models with the present spectrum. We conclude that the O
K-shell column densities indeed did vary between observations.
This can be seen even in our Fig.~\ref{plotone} where features of
O$^{+6}$ and O$^{+7}$, marked by labels, are clearly deeper in the
present spectrum, even though the soft excess continuum is higher.
Conversely, for all non-oxygen ions, we find that the present
ionic column densities obtained from fitting the \hetgs\ spectrum
actually reproduce the \rgs\ spectrum fairly well. Other than O,
we have no good explanation for the discrepancies with
\citet{sako01}.

Interestingly, we find a relatively high column density of neutral
Fe which may not be part of the outflow, as manifested by the
L$\alpha$ absorption \uta~at 17.53 \AA~(see Fig.~\ref{plotthree}).
If we use the solar Fe/H abundance ratio \citep{as05}, we obtain a
neutral gas column density of $N_{H}\sim3.5\times10^{21}$~\cmsq,
much higher than the galactic column $N_{H}\sim10^{20}$~\cmsq.
This may indicate the presence of neutral gas or dust in the host
galaxy reminiscent of the \citet{wills92} finding.

%In any case the densities found at \citet{sako01} are not enough
%to explain some absorption features like O$^{+6}$ He$\beta$ or
%O$^{+7}$ Ly$\gamma$.

\subsection{Errors Determination}
The column density errors quoted in Table~\ref{tab1} were
calculated by keeping an individual ionic column density fixed
while performing a minimum -- $\chi ^2$ fit with all other
parameters, and then requiring that $\Delta \chi ^2 = 1$ for
variations of that individual ionic column density, while all
other parameters remain fixed. The more statistically rigorous
method would be to require $\Delta \chi ^2 = 1$ following a refit
of the data. However, since we supplement $\chi ^2$ fitting with a
more controlled manual fit to the strongest lines, leading to a
best fit that is not strictly the $\chi ^2$ minimum, this method
which involves automatic $\chi ^2$ minimization is inapplicable
here. Nevertheless, our simpler method gives reliable upper limits
to the errors.

For the cases in which the relative error was less than 10\%, we
assume an error of 10\%.  The reason is that even for the most
prominent lines, we do not trust the column density to better than
10\%. Conversely, when the relative error was more than 200\%, we
quote a nominal error of 200\%. In these cases we believe the
errors are not statistical.

% Note that
%usually for strong, nearly saturated absorption lines,
%statistically determined errors come out highly asymmetrical.
%Raising the column density affects a saturated line (and thus
%$\chi^2$) very little, while on the other hand, the fit is rather
%sensitive to a decrease in the column density, especially for
%those ions with noticeable edges like O K-shell or Ne K-shell
%because their edges affect a much larger region of the spectrum
%and thus have more effect on $\chi ^2$.

% For the analysis in the next section,
%symmetrical errors were more desirable. Thus, we used the
%geometrically averaged error:

%\begin{equation}
% err_{symmetric} = \sqrt{err_{up}\cdot err_{down}}.
%\end{equation}

In \S \ref{sec:$AMD$} we need to use symmetrical errors. In order
to get a reliable symmetrical error estimate, we introduced one
more adjustment. We define a geometrically averaged error :

\begin{equation}
 err_{symmetric} = \sqrt{err_{up}\cdot err_{down}}.
\end{equation}

%For the cases in which $err_{symmetric} < 0.2N_{ion}$, we assume
%an error of $0.2N_{ion}$. Conversely, when $err_{symmetric}
%> 0.9N_{ion}$, we quote a nominal error of $0.9N_{ion}$.
%The reason for these cutoffs is the crude assumption that even for
%the most prominent lines, we do not trust the column density to
%better than 20\%, while for an identified line, we assume that at
%least 10\% of $N_{ion}$ is present at a 1$\sigma$ confidence
%level.

\noindent We use geometrically averaged rather than arithmetically
averaged errors, since the geometrical average gives an average in
order of magnitude between the two errors, which is the more
appropriate value to be applied as a symmetrical error. For the
full $AMD$ analysis (presented in \S \ref{sec:$AMD$}), we use the
true asymmetrical errors in the manner explained below.

%Other ions that are not conclusively
%identified are not included in Table~1.

\section{ABSORPTION MEASURE DISTRIBUTION}
\subsection{Method}
\label{sec:$AMD$}

The large range of ionization states present in the absorber
strongly suggests that the absorption arises from gas that is
distributed over a wide range of ionization parameter. The total
hydrogen column density $N_H$ along the line of sight can
therefore be expressed as an integral over its distribution in
$\log \xi$. We call this continuous distribution the Absorption
Measure Distribution ($AMD$ by analogy to the Emission Measure
Distribution in emission-line spectra):

\begin{equation}
AMD=dN_H/d(\log \xi)
\end{equation}

and

\begin{equation}
N_H = \int AMD~d(\log \xi)
\end{equation}

\noindent The relation between the ionic column densities
$N_{ion}$ and the $AMD$ is then expressed as:

\begin{equation}
N_{ion} = A_z\int\frac{dN_H}{d(\log\xi)}f_{ion}(\log
\xi)d(\log\xi)
 \label{eqAMD}
\end{equation}

\noindent where $N_{ion}$ is the measured ion column density,
$A_z$ is the element abundance with respect to hydrogen assumed to
be constant throughout the absorber, and $f_{ion}(\log\xi)$ is the
fractional ion abundance with respect to the total abundance of
its element. Here, we aim at recovering the $AMD$ for \iras.

\begin{figure}
%  \plotone{plotfour.ps}
% \centerline{\psfig{figure=NeIX.eps,height=7in}}
%
\centerline{\includegraphics[width=13cm,angle=90]{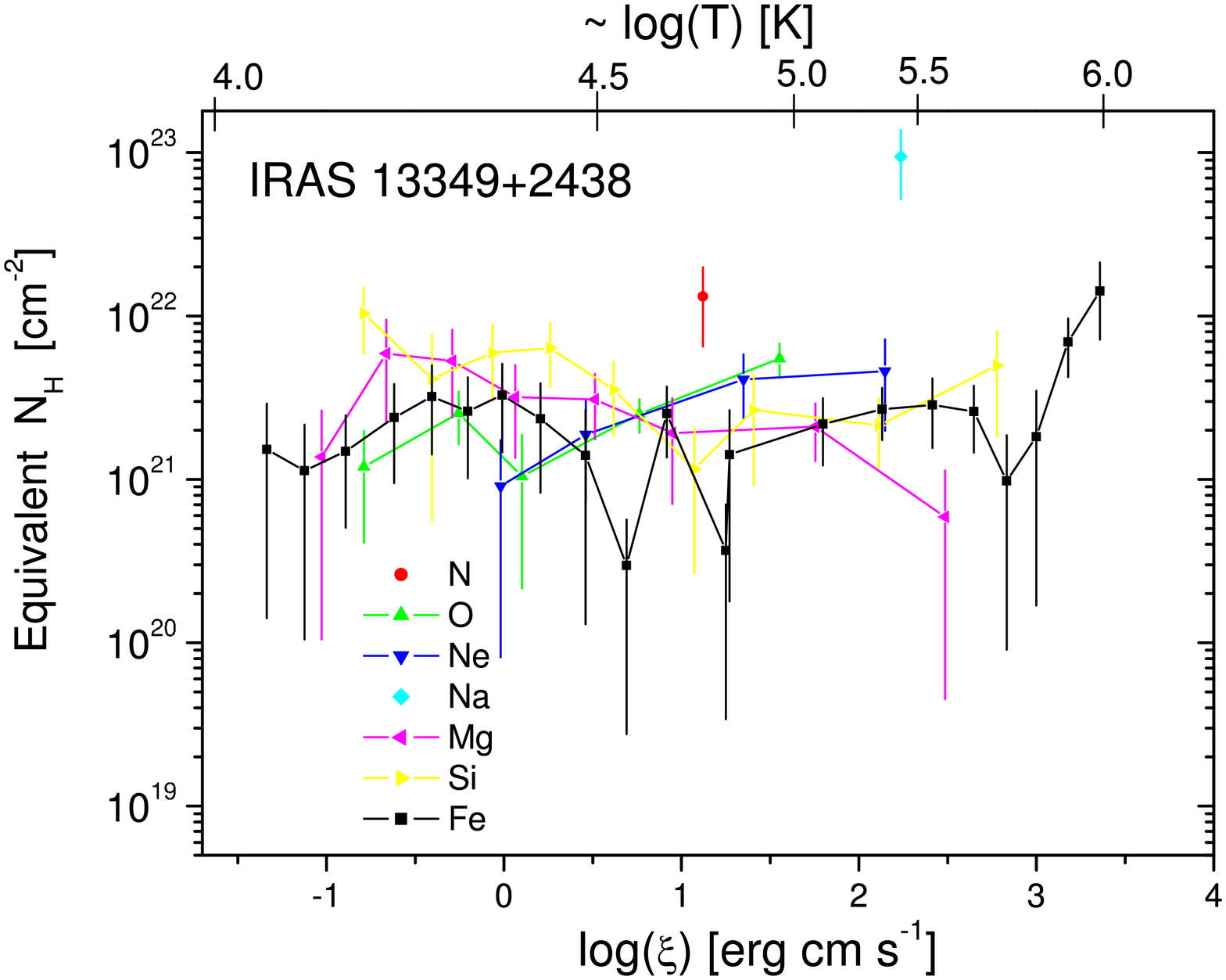}}
  \caption{Equivalent $N_H$ distribution (eq.~\ref{eqNH}) obtained for the \iras\ outflow
  assuming ions form at $\xi _{max}$ and assuming solar abundances \citep{as05}.
  Lines are drawn between data points just to guide the eye.
  Vertical offsets between elements indicate deviations from solar abundances.
  The corresponding temperature scale, obtained from the \xstar\ computation
  is shown at the top of the figure.}
   \label{plotfour}
\end{figure}

As an initial approximation to be relaxed later, let us assume
that each ion forms exclusively at the ionization parameter $\xi
_{max}$ where its fractional abundance peaks. Furthermore, if
solar abundances $A_{Z_\odot}$ are assumed, the equivalent
hydrogen column density can be calculated separately from each ion
using the relation:

\begin{equation}
N_H \simeq\ \frac{N_{ion}}{f_{ion}(\xi _{max}) A_{Z_\odot}}
 \label{eqNH}
\end{equation}

\noindent and placed at the position of $\xi _{max}$ on an
$N_H(\log \xi)$ plot. For this we employed the \xstar\ code
\citep{kal95} version 2.1kn3 to calculate $f_{ion}(\log\xi)$ using
the continuum derived in \S \ref{sec:continuum} extrapolated to
the range of 1 -- 1000 Rydberg. The results for \iras\ using solar
abundances from \citet{as05} are presented in Fig.~\ref{plotfour}.
The ions pertaining to each element in the figure are connected by
straight lines to guide the eye. The fact that the connecting
lines of Fe, Si, Mg, O and Ne largely overlap on this plot implies
that the relative abundances of these elements do not deviate much
from the solar values. On the other hand, the relative abundance
of N and Na appears to be significantly above solar, although
there is only one N and one Na ion to show this result. The errors
in Fig.~\ref{plotfour} are those propagated from the $N_{ion}$
symmetrical uncertainties as explained in the previous section.

The distribution presented in Fig.~\ref{plotfour} is only a first
approximation for the actual $AMD$. For the real $AMD$, we need to
find a distribution $d N_H / d(\mathrm{\log}\xi)$ that after
integration (eq.~\ref{eqAMD}) will produce all of the measured
ionic column densities (Table~\ref{tab1}). In fitting for the
$AMD$ one must take into account the full dependence of $f_{ion}$
on $\xi$.
% We employ the \xstar\ code
%\citep{kal95} version 2.1kn3 to calculate $f_{ion}(\log\xi)$ using
%the continuum derived in \S \ref{sec:continuum} extrapolated to
%the range of 1 -- 1000 Rydberg.
We assume all charge states see the same ionizing spectrum. This
is justified by the absence of significant bound-free absorption
edges in the spectrum. All elements are expected to reflect the
same $AMD$ distribution, due to the assumption that they all
reside in the same gas. Iron however, has a special role as it
covers almost five orders of magnitude in $\xi$, more than any
other element. Nonetheless, in order to improve the $AMD$ fit, we
can use other elements as well. In order to incorporate the non-Fe
ionic column densities without assuming anything about the
elemental abundances, for non-Fe ions we use column density
ratios. Using a reference ion for each element (the one with the
smallest symmetrical errors), these ratios can be written as:

\begin{equation}
\frac{N_{ion}}{N_{ion-reference}} = \frac{A_z\int\frac{d
N_H}{d(\log\xi)}f_{ion}(\log \xi)d(\log\xi)}{A_z\int\frac{d
N_H}{d(\log\xi)}f_{ion-reference}(\log \xi)d(\log\xi)}
\label{eqratio}
\end{equation}

Since there is no {\it a-priori} physical argument for a specific
functional form for the $AMD$, we choose to use a simple staircase
function (Fig.~\ref{plotfive}). The main advantage of this form is
the well-localized easily-calculated errors. A similar approach
was used for emission by \citet{nordon06}.

The fit is performed with our Monte-Carlo evolution program in
which the only free parameters are the values of the $AMD$ in each
log$\xi$ bin. The program receives starting values for each bin,
then generates 5000 other $AMD$ solutions randomly in the
multi-dimensional region around the starting point. For each
solution, $\chi ^2$ is calculated and the lowest-$\chi ^2$
solution is subsequently considered as the new starting point.
This process is repeated until $\chi ^2$ can not improve any more.
In our tests, the program always converged to the same best-$AMD$
solution for arbitrary starting values within five orders of
magnitude (higher or lower) of the final result. This gives us
confidence that the fit finds the global minimum. As this is an
ill-defined inversion problem, the $AMD$ binning is driven by the
correlated errors. At first, we use 10 bins in the range of
log$\xi$ = --1.25 to 3.75 (c.g.s.), assuming the $AMD$ is zero
elsewhere. We then combine bins so that the range and number of
bins are optimized so that $AMD$ bin values are inconsistent with
zero (to within 1$\sigma$), except in broad regions of $\xi$ in
which the $AMD$ is persistently zero. The final number of bins
(i.e., the $AMD$ resolution) depends on the specific absorber and
on the quality of the data, which can limit the bin size.

% The objective is to obtain $AMD$ bins whose values are
%inconsistent with zero to within the errors. But if there are bins
%that their value is consistent with zero and even after adding
%them up with more stable bins, their value will still be
%consistent with zero within the errors, we don't add them up and
%we find only an upward error.
The $AMD$ fit is obtained by the $\chi ^2$ minimalization
technique with respect to all of the measured Fe $N_{ion}$ values
(eq.~\ref{eqAMD}) and $N_{ion}$ ratios of at least one other
element (eq.~\ref{eqratio}). In the calculation of $\chi ^2$, the
real asymmetrical errors are used so that when the best-fit $AMD$
overestimates an ionic column density, the upper limit error is
used and vice versa. For non-Fe ions, the same rule is applied to
$N_{ion}$, while symmetrical errors are used for
$N_{ion-reference}$ (eq.~\ref{eqratio}).

The $AMD$ errors are calculated in each bin including correlations
between bins. The $AMD$ in a given bin is varied from its best-fit
value and the whole distribution is refitted. This procedure is
repeated until $\Delta \chi ^2= 1$. The fact that changes in the
$AMD$ in one bin can be compensated by varying the $AMD$ in other
bins dominates the $AMD$ uncertainties. This is what limits the
number of bins and the $AMD$ resolution. $AMD$ in neighboring,
excessively narrow bins can not be distinguished by the data,
i.e., different distributions produce the measured $N_{ion}$
values to within the errors. A meaningful quantity is the integral
of the $AMD$ up to $\xi$ (as a function of $\xi$) since during
integration correlated errors cancel out, which explains the small
errors on the integrated column density (Fig.~\ref{plotfive},
bottom panel).

%bin raised (or lowered) and fixed ,then a new fit (via Monte
%Carlo) was made . when the raise (or lower) was so high that the
%new fit`s $\chi^2$ was larger by a unity - we took that value as
%an upper (or lower) limit of the bin in 1$\sigma$ confidence
%level. Until that value all the other bins could compensate for
%that bin`s change , so when we took a large number of bins (like
%20) then the lower limit on most of them was consistent with zero,
%which means that the whole bin could be compensated by neighbor
%bins and as a result - we cant really say anything about that bin.
%In order to obtain the covariance errors between the bins (needed
%for the integrated column density - see figure 5),for each bin we
%took 68 values for that bin which are inside the error bars
%(1$\sigma$) and another 32 values which are between
%1$\sigma$-2$\sigma$, for each one of these values we made a best
%"new" fit and using the equation for covariance -
%\begin{equation}
%cov(x,y)~=~\frac{1}{N}\sum(x_i-\bar{x})(y_i-\bar{y})
%\end{equation}
% where $x_i$
%and y$_i$ are the values of the bins in the "new" fit and
%$\bar{x}$ and $\bar{y}$ are the values in the best fit.

The next step is to estimate the element abundances relative to
Fe. This is done by assuming that the other elements are
distributed in log$\xi$ similar to Fe, or in other words, that the
{\it shape} of the $AMD$ (e.g., Fig.~\ref{plotfive}) is manifested
in the column density distribution of each and every element. This
is what one would expect for a chemically uniform absorber.
Indeed, the overlap of elements in Fig.~\ref{plotfour} seems to
support this assumption.
% As a sanity check, we produced an $AMD$ plot from the
%measured column densities of Si ions and obtained a very similar
%distribution to that of Fig.~9.
In order to obtain the abundances, we used the $AMD$ distribution,
by varying the $AMD$ normalization to best-fit the column
densities of each ion. This single-parameter fit yields for each
ion the relative abundance $A_Z/A_{Fe}$. The errors on these
abundances can be calculated from the $\Delta \chi^2= 1$
requirement, again using the asymmetrical column density errors,
but also the uncertainties on the $AMD$ distribution itself. We
then average this result for all ions from the same element to
obtain our best abundance estimate. The errors on the abundances
are calculated by averaging the upward and downward confidence
limits separately.

%and with respect to the solar abundances \citep{as05}, i.e.,
%$(A_Z/A_{Fe})/(A_Z/A_{Fe})_\odot$.

\subsection{Results}
\begin{figure}
%  \plotone{plotfive.ps}
% \centerline{\psfig{figure=NeIX.eps,height=7in}}
%
\centerline{\includegraphics[width=13cm,angle=0]{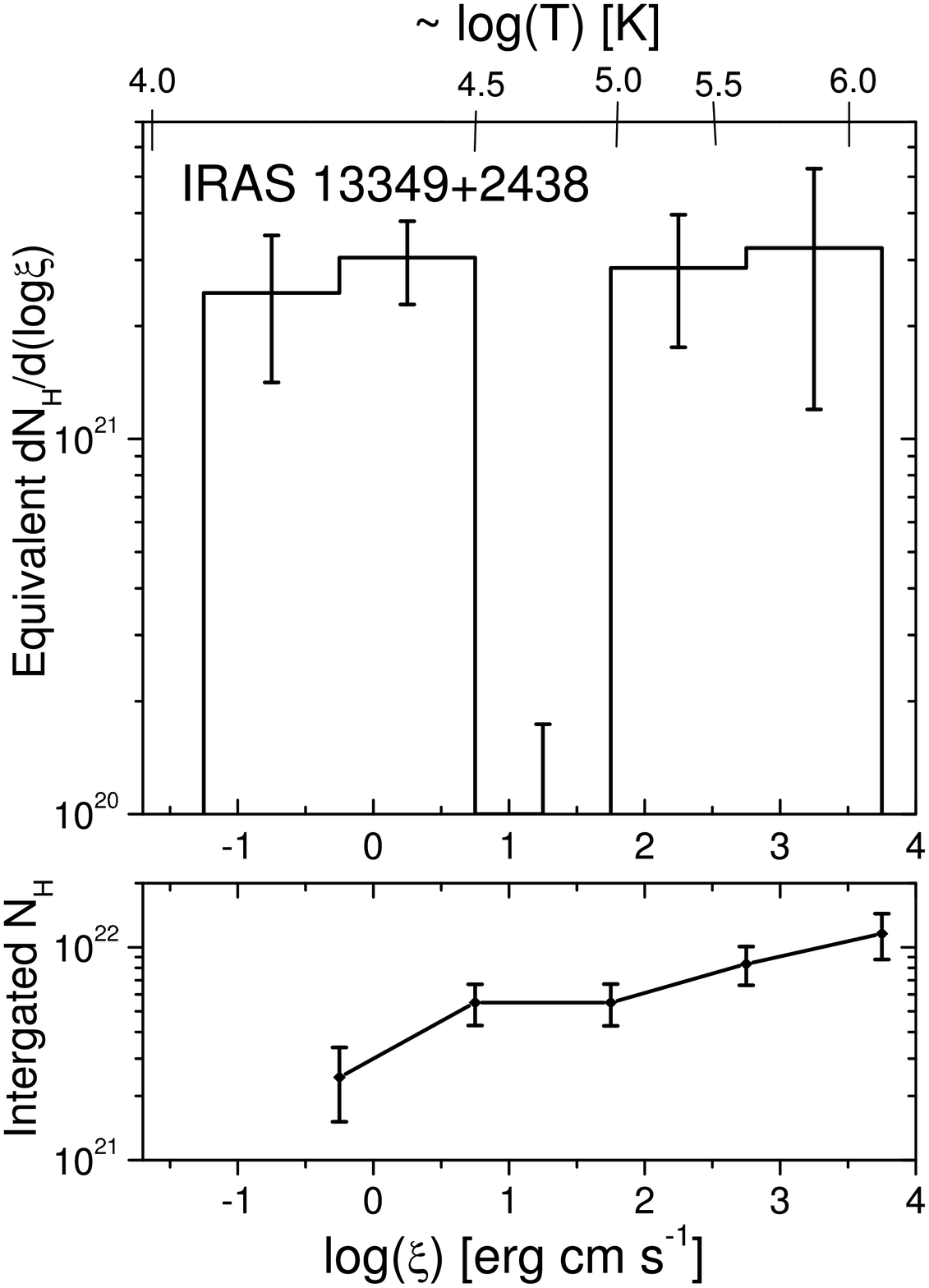}}
  \caption{$AMD$ of \iras~scaled by the solar Fe/H abundance
  2.82$\times10^{-5}$. The corresponding temperature scale, obtained from the \xstar\ computation
  is shown at the top of the figure. The third bin value is zero,
  so only an upper limit error was calculated
  Note the minimum at $0.75 < \log\xi < 1.75$ (c.g.s. units) corresponding to $4.5 < \log T <5$ (K).
  The integrated column density up to $\xi$ is shown in the
   lower panel yielding a total of $N_H = (1.2\pm0.3)\times10^{22}$~\cmsq.}

   \label{plotfive}
\end{figure}

The best-fit $AMD$ for the absorber in \iras\ is presented in
Fig.~\ref{plotfive}. The integrated column density for the
absorber of \iras\ is also presented in Fig.~\ref{plotfive} in the
lower panel. This $AMD$ was obtained using 21 ions of Fe and 9
ions of Si. The integrated $AMD$ of the absorber in \iras\
(Fig.~\ref{plotfive}) gives a total column density of
(1.2$\pm$0.3)$\times10^{22}$~\cmsq\ compared with
(1--4)$\times10^{22}$~\cmsq\ in \citet{sako01}. The $AMD$ features
a statistically significant minimum at $0.75 < \log\xi < 1.75$
(c.g.s. units), which corresponds to $4.5 < \log T <5$ (K) and
which was noted also by \citet{sako01}. This minimum is a
manifestation of the relatively low ionic column densities
observed for the ions Fe$^{+11}$--~Fe$^{+15}$ seen in
Fig.~\ref{plotfour} and in Table~\ref{tab1}.

\begin{figure}
%  \plotone{plotsix.ps}
% \centerline{\psfig{figure=NeIX.eps,height=7in}}
%
\centerline{\includegraphics[width=13cm,angle=90]{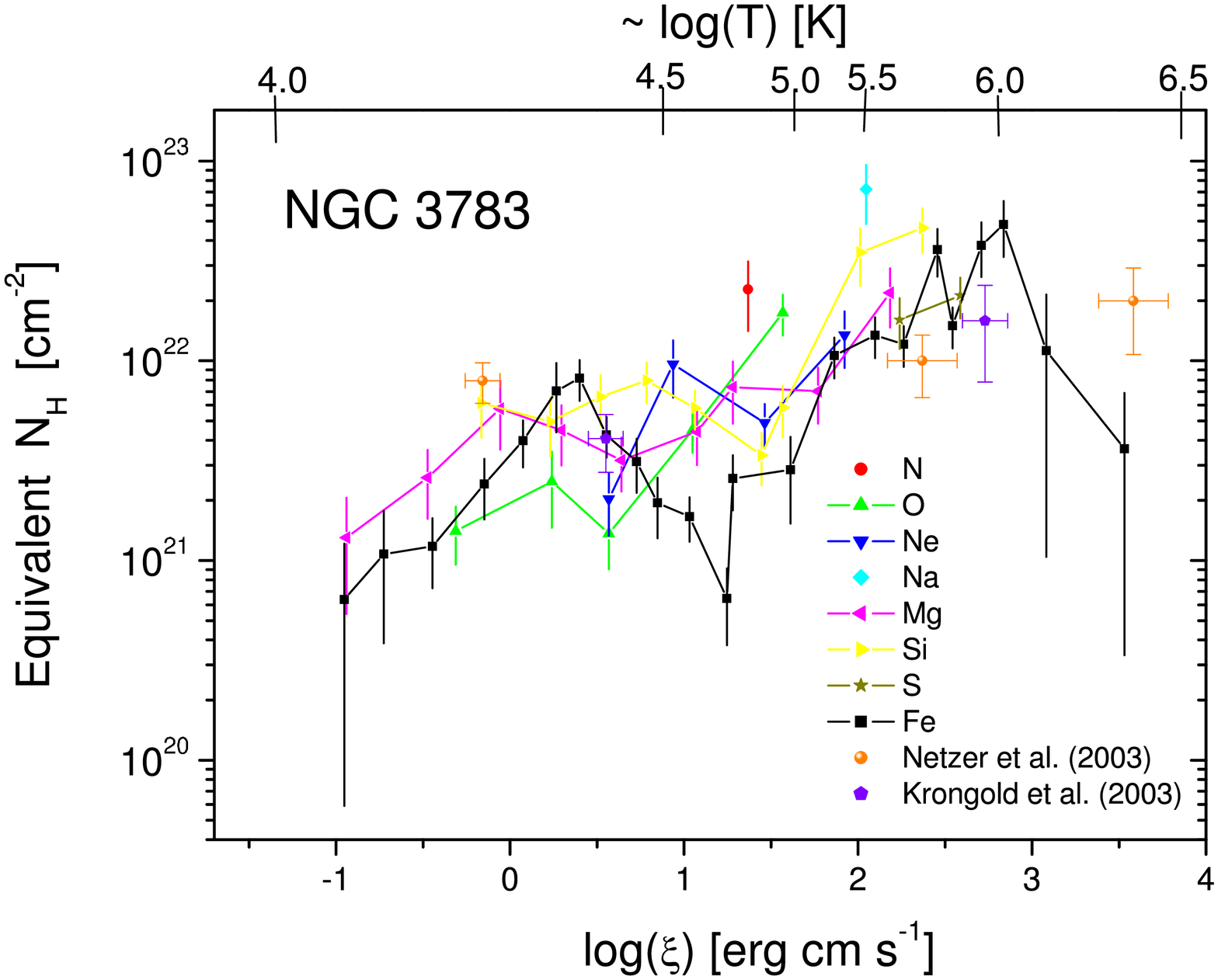}}
  \caption{Equivalent $N_H$ distribution (eq.~\ref{eqNH}) obtained for the \ngc\ outflow
  assuming ions form at $\xi _{max}$ and assuming solar abundances \citep{as05}.
  Lines are drawn between data points just to guide the eye.
  Vertical offsets between elements indicate deviations from solar abundances.
  The corresponding temperature scale, obtained from the \xstar\ computation
  is shown at the top of the figure.
  The \citet{netzer03} three component model results and the \citet{krongold03}
  two component model results are plotted for comparison.}

   \label{plotsix}
\end{figure}

\begin{figure}
%  \plotone{ngcdint.ps}
% \centerline{\psfig{figure=NeIX.eps,height=7in}}
%
\centerline{\includegraphics[width=11cm,angle=0]{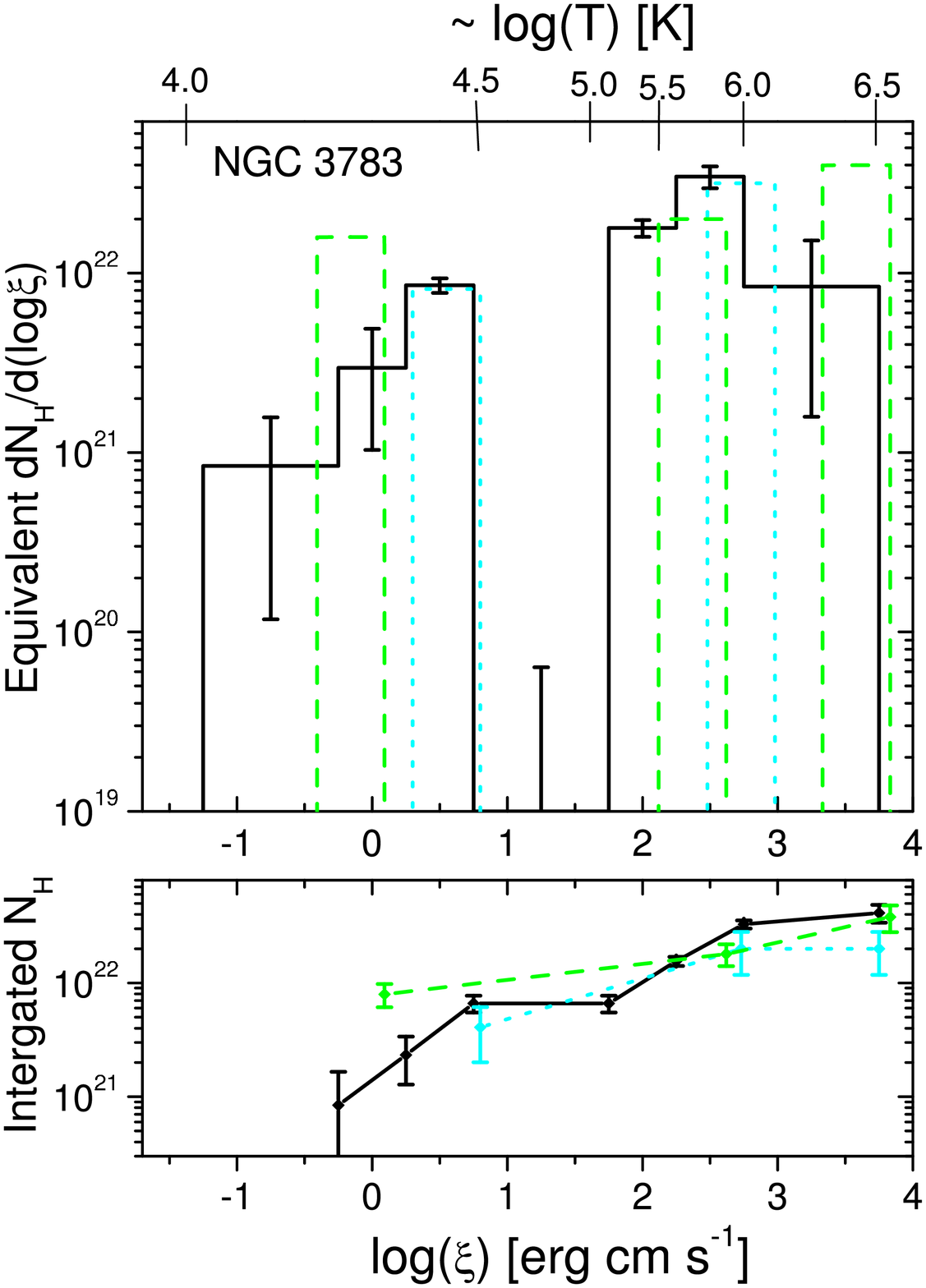}}
  \caption{$AMD$ of \ngc~scaled by the solar Fe/H abundance of $2.8\times10^{-5}$. The corresponding temperature scale, obtained from the \xstar\ computation
  is shown at the top of the figure.
  Note the minimum at $0.75 < \log\xi < 1.75$ (c.g.s. units) corresponding to $4.5 < \log T <5$ (K),
  a behavior similar to that of
  \iras, but better constrained here.
  The fourth bin value is zero so only an upper limit error is calculated.
  The integrated column density up to $\xi$ is shown in the lower panel
  yielding a total of $N_H = (4.1\pm0.7)\times10^{22}$~\cmsq.
  Green dashed and cyan dotted lines represent the three- and two- components results, respectively,
   of \citet{netzer03} and of  \citet{krongold03} where each component is broadened arbitrarily to
   $\Delta$log$\xi = 0.5$ (c.g.s. units) bins.}
   \label{plotseven}
\end{figure}

In order to check that the observed minimum in the $AMD$ of \iras\
is not a result of the S/N, we want to apply the same method to
another ionized absorber with a wide range of ionization. The
natural choice is the 900~ks \hetgs\ spectrum of \ngc. We use the
spectrum from \citet[Fig. 5]{kaspi02}, and the same method
described above. The results are presented in Figs.~\ref{plotsix}
and \ref{plotseven} in the same format as for \iras\
(Figs.~\ref{plotfour} and \ref{plotfive}). The $AMD$ of \ngc\ was
obtained using 23 ions of Fe and 4 ions of Ne. Already in
Fig.~\ref{plotsix}, low column densities for the
Fe$^{+12}$--~Fe$^{+16}$ and Si$^{+9}$--~Si$^{+11}$ charge states
can be seen around log$\xi \eqsim$ 1--2 (c.g.s. units). Indeed, in
the staircase $AMD$ reconstruction for \ngc\ presented in
Fig.~\ref{plotseven}, we find a minimum at $0.75 < \log \xi <1.75$
(c.g.s. units), which corresponds to $4.5 < \log T < 5$ (K). This
minimum is surprisingly similar to that of \iras, but is more
statistically significant and thus better constrained in $\xi$. We
note that a similar minimum has been recently observed in {\small
NGC~7469} as well \citep{blustin07}.

In Figs.~\ref{plotsix} and \ref{plotseven}, we include comparisons
with the two-component model of \citet{krongold03} and the
three-component model of \citet{netzer03} for the same \ngc\
\hetgs\ data set. The ionization components of these papers are
placed on the plot according to $T$ rather than $\xi$ as different
models produce different relations between temperature and $\xi$.
%In Fig.~11 we present
%these models by dashed-dotted cyan \citep{krongold03} and by
%dotted green \citep{netzer03} lines.
The bin size, is set arbitrarily to $\Delta \log \xi = 0.5$
(c.g.s. units). The column density of each component in those
papers then determines the area (height) of its $AMD$ contribution
in Fig.~\ref{plotseven}. It can be seen that all models avoid the
region of the $AMD$ minimum between log$\xi \simeq$~ 1 and 2
(c.g.s. units). On the other hand, only the present method, that
allows for a complete distribution, accounts for the full range of
absorbing gas from log$\xi$ = $-$1.25 to log$\xi$ = 3.75 (c.g.s.
units).
%The wide spread of $AMD$ as a function of $\xi$ results in a higher
% integrated column density.
We obtain a total (integrated) $N_H$ value of
(4.1$\pm$0.7)$\times10^{22}$~\cmsq\ compared with
(3.8$\pm$1.0)$\times10^{22}$~\cmsq\ and
(2.0$\pm$0.8)$\times10^{22}$~\cmsq\ obtained by \citet{netzer03}
and by \citet{krongold03}, respectively. Note, that we use
slightly different solar abundances than the other authors do.

Finally, the results for the chemical abundances obtained for both
\iras\ and \ngc\ using the method described in the previous
section are listed in Table~\ref{tab2}. We include in the table
only elements with at least two significant ion detections. In
both \agn s the Ne/Fe, Mg/Fe and O/Fe abundance ratios are
slightly above or at about their solar value. The Si/Fe ratio is
above solar. The S/Fe ratio can be measured only for \ngc\ where
it is at its solar value. In general we can say that in both \agn
s the metal abundances are approximately consistent with solar
values.
%We note that
%the oxygen absorption has appeared to change compared to the early
%\rgs\ observation \citep{sako01}, while Fe , N and Ne absorption,
%as far as we can tell, has remained the same.

\subsection{Discussion: Observing Thermal Instability}
The extent to which we can describe the low ionization region of
the $AMD$  strongly depends on the photoionization balance, which
is highly uncertain at low-$\xi$ for two main reasons. First, the
uncertainties in the dielectronic recombination rates of the low
charge states of Fe at low (photoionized) temperatures are well
known to be severely underestimated in all codes including \xstar;
see \citet{netzer04} and recently \citet{badnell06}. Correcting
these would shift the ionization balance at a given $\xi$ value
towards lower charge states, which would result in a more compact
$AMD$ not extending as far on the low-$\xi$ end. The other effect
that could drastically change our low-$\xi$ $AMD$ is
photoionization of the low charge states by the EUV and UV
continuum, which is only poorly constrained. For example, a softer
continuum than we assumed here (see \S \ref{sec:continuum}), would
allow for more highly ionized M-shell charge states at a given
$\xi$ value. Consequently, it is important to understand that the
exact position of the low-$\xi$ peak of the $AMD$ for both \iras\
and \ngc\ is only poorly constrained, and will depend on the
combined correction for the DR rates and for the soft continuum.
Conversely, the total column density in this component as well as
the existence of a minimum in the $AMD$ are robust.

One explanation that comes to mind for the minimum observed in the
$AMD$ of both \iras\ and \ngc\ is thermal instability. If
photoionized gas is thermally unstable at these temperatures, it
would provide a natural explanation for the absence of absorption
and for the low column densities of the charge states that form
primarily at these temperatures. Note that our $AMD$ binning is
driven by the requirement for meaningful errors. Consequently, we
can not determine where the alleged instability occurs to better
accuracy than the $\xi$ and $T$ regimes given above, as narrower
bins will inevitably result in excessive $AMD$ uncertainties. The
resulting $AMD$ distributions presented in Figs.~\ref{plotfive}
and~\ref{plotseven} along with their similar minima are
reminiscent of what is widely referred to in the literature as the
two-phase models of gas in photoionization equilibrium
\citep{krolik81}. These models have been invoked for \agn\ most
importantly to allow for the confinement of low-$T$ high-density
clouds by high-$T$ diffused gas. It is important to stress that
the currently derived $AMD$ is not a model. It represents a direct
measurement of the ionization structure of the absorber that is
more complete and more detailed than a multi iso-$\xi$ fit. To
that end, the observed $AMD$ minima provide direct evidence for a
thermal unstable temperature region that can not be obtained from
iso-$\xi$ fits.

The observed minimum or two-phase structure can also be ascribed
to two geometrically distinct regions along the line of sight, a
high ionization region and a low ionization region, both which
have their own gaussian like $AMD$ distribution. However, there
are good reasons to believe that gas at $4.5 < \log T < 5$ (K)
would be unstable as the cooling function $\Lambda(T)$ is
generally decreasing in this temperature regime \citep{krolik81}.
Additionally, the model of \citet[Fig.~12]{netzer03} for \ngc\
shows that this region is marginally unstable. The model of
\citet{goncalves06} for \ngc\ avoids temperature zones, which
could be a result of thermal instability. Interestingly, the
avoided region at $4.5 \lesssim \log T \lesssim 5$ (K) is
consistent with the present measurement. However, there is another
region of avoidance in that model at $5.3 \lesssim \log T \lesssim
6$ (K) that our measurement does not show. The model of
\citet{krongold03} has a very wide unstable region at $5.3
\lesssim \log T \lesssim 7.5$. However, both \citet{netzer03} and
\citet{krongold03} find an absorption component in this
temperature regime, as we do. As the regions of instability depend
strongly on unobserved regions of the ionizing spectra, the
theoretical stability analysis models have to be treated with
caution.

%One element that is appreciably overabundant with respect to the
%Sun in both sources is nitrogen as also seen in NGC 1068
% \citet{brinkman02}. This result has to be taken with caution as we
% observe absorption by only one ion of N (N$^{+6}$) in the spectrum.
%Moreover, this ion forms preferably in what appears to be the
%thermally unstable region.

\section{CONCLUSIONS}
\label{sec:concl}

We have analyzed the thermal and chemical structure of the ionized
outflow in the quasar \iras. Using a new $AMD$ reconstruction
method, we measure the distribution of column density in the
outflow as a function of $\xi$. We find a double-peaked
distribution with a significant minimum at $0.75 < \log \xi <
1.75$ (c.g.s. units), which corresponds to temperatures of $4.5 <
\log T < 5$ (K). Using a comparison method of the $AMD$ derived
for different elements we are able to estimate the relative
chemical abundances in the outflow.
%We find that the Ne/Fe ratio is subsolar, while the N/Fe is supersolar. Other element
We find that all of the abundances relative to Fe are more or less
solar. For comparison, we applied the same analysis to the
excellent-quality \x\ spectrum of \ngc. We find a similar minimum
in the $AMD$ and rather similar abundances suggesting perhaps that
these could be common features in \agn\ outflows. We believe this
minimum is due to thermal instability.

\acknowledgments This research was supported by grant \#28/03 from
the Israel Science Foundation, by a grant from the Asher Space
Research Institute at the Technion. S.K. acknowledges support of
the Zeff fellowship at the Technion.

\clearpage
\begin{deluxetable}{lccccc}
\tabletypesize{\normalsize} \tablecolumns{5} \tablewidth{0pt}
\tablecaption{Best-fit
 column densities for ions detected in the 2004 \hetgs\ spectrum
 of \iras\ and comparison to the 2000 \rgs\ observation \citep{sako01}.
 \label{tab1}} \tablehead{
   \colhead{Ion} &
   \colhead{\hetgs\  } &
   \colhead{\rgs\ } &
   \colhead{Ion} &
   \colhead{\hetgs\   } &
   \colhead{\rgs\  } \\
   \colhead{} &
   \colhead{Column Density } &
   \colhead{Column Density} &
   \colhead{} &
   \colhead{Column Density } &
   \colhead{Column Density } \\
   \colhead{} &
   \colhead{(10$^{16}$~cm$^{-2}$)} &
   \colhead{(10$^{16}$~cm$^{-2}$)} &
   \colhead{} &
   \colhead{(10$^{16}$~cm$^{-2}$)} &
   \colhead{(10$^{16}$~cm$^{-2}$)}
}
  \startdata
N$^{+6}$    &   $45_{-23}^{+21}$    &   $13_{-4}^{+8}$  &   Fe$^{+0}$   &   $9.9_{-5.2}^{+3.0}$ &   \nodata \\
\cline{1-3}
O$^{+3}$    &   $30_{-18}^{+21}$    &   \nodata     &   Fe$^{+3}$   &   $1.3_{-1.3}^{+2.2}$ &   \nodata \\
O$^{+4}$    &   $40_{-36}^{+5.0}$ &   \nodata         &   Fe$^{+4}$   &   $1.0_{-1.0}^{+1.6}$ &   \nodata \\
O$^{+5}$    &   $15_{-3.3}^{+30}$   &   \nodata         &   Fe$^{+5}$   &   $1.9_{-1.9}^{+0.8}$ &   \nodata \\
O$^{+6}$    &   $78_{-8.0}^{+14}$ &   $3.7_{-1.8}^{+2.7}$     &   Fe$^{+6}$   &   $2.0_{-2.0}^{+0.7}$ &   $1.5_{-1.3}^{+1.5}$ \\
O$^{+7}$    &   $140_{-39}^{+21}$   &   $9.5_{-4.7}^{+8.7}$         &   Fe$^{+7}$   &   $2.7_{-2.7}^{+0.8}$ &   $4.6_{-1.3}^{+1.5}$ \\
\cline{1-3}
Ne$^{+6}$   &   $2.7_{-2.1}^{+3.3}$ &   \nodata     &   Fe$^{+8}$   &   $1.8_{-1.3}^{+0.9}$ &   $0.9_{-0.9}^{+1.2}$ \\
Ne$^{+7}$   &   $5.5_{-3.0}^{+3.6}$ &   \nodata &   Fe$^{+9}$   &   $2.5_{-2.3}^{+0.8}$ &   $2.4_{-1.1}^{+1.3}$ \\
Ne$^{+8}$   &   $22_{-2.2}^{+35}$ &   $120_{-60}^{+50}$   &   Fe$^{+10}$  &   $2.0_{-1.8}^{+0.9}$ &   $1.9_{-0.9}^{+1.1}$ \\
Ne$^{+9}$   &   $16_{-6.7}^{+12}$   &   $49_{-32}^{+100}$   &   Fe$^{+11}$  &   $1.4_{-1.4}^{+1.2}$ &   $6.4_{-4.4}^{+10}$ \\
\cline{1-3}
Na$^{+10}$  &   $7.0_{-3.2}^{+3.0}$ &   \nodata     &   Fe$^{+12}$  &   $0.2_{-0.2}^{+0.4}$ &   \nodata \\
\cline{1-3}
Mg$^{+4}$   &   $2.0_{-2.0}^{+4.0}$ &   \nodata     &   Fe$^{+13}$  &   $2.0_{-2.0}^{+0.4}$ &   \nodata \\
Mg$^{+5}$   &   $8.0_{-5.0}^{+4.3}$ &   \nodata     &   Fe$^{+14}$  &   $0.4_{-0.4}^{+0.8}$ &   \nodata \\
Mg$^{+6}$   &   $6.2_{-5.5}^{+1.8}$ &   \nodata     &   Fe$^{+15}$  &   $0.8_{-0.8}^{+0.6}$ &   \nodata \\
Mg$^{+7}$   &   $4.6_{-3.8}^{+1.6}$ &   \nodata     &   Fe$^{+16}$  &   $2.7_{-1.1}^{+1.2}$ &   $17_{-12}^{+16}$ \\
Mg$^{+8}$   &   $4.4_{-2.3}^{+1.2}$ &   \nodata     &   Fe$^{+17}$  &   $2.7_{-2.0}^{+0.4}$ &   $63_{-19}^{+24}$ \\
Mg$^{+9}$   &   $2.8_{-2.0}^{+1.4}$ &   \nodata     &   Fe$^{+18}$  &   $2.8_{-1.9}^{+0.8}$ &   $97_{-39}^{+48}$ \\
Mg$^{+10}$  &   $5.5_{-1.6}^{+2.0}$ &   \nodata     &   Fe$^{+19}$  &   $2.5_{-1.6}^{+0.7}$ &   $30_{-20}^{+28}$ \\
Mg$^{+11}$  &   $1.0_{-1.0}^{+2.0}$ &   \nodata     &   Fe$^{+20}$  &   $0.5_{-0.5}^{+0.9}$ &   \nodata \\
\cline{1-3}
Si$^{+5}$   &   $15_{-15}^{+2.7}$   &   \nodata     &   Fe$^{+21}$  &   $1.6_{-1.0}^{+2.6}$ &   \nodata \\
Si$^{+6}$   &   $5.0_{-5.0}^{+3.7}$ &   \nodata     &   Fe$^{+22}$  &   $5.0_{-5.0}^{+0.7}$ &   \nodata \\
Si$^{+7}$   &   $7.0_{-3.4}^{+3.3}$ &   \nodata     &   Fe$^{+23}$  &   $8.0_{-4.0}^{+3.7}$ &   \nodata \\
\cline{4-6}
Si$^{+8}$   &   $7.0_{-3.2}^{+2.6}$ &   \nodata     &   Ni$^{+18}$  &   $0.6_{-0.4}^{+0.7}$ &   \nodata \\
Si$^{+9}$   &   $4.9_{-2.0}^{+2.6}$ &   \nodata     &               &   \\
Si$^{+10}$  &   $1.7_{-1.3}^{+1.3}$ &   \nodata     &               &   \\
Si$^{+11}$  &   $3.0_{-1.3}^{+2.9}$ &   \nodata     &               &   \\
Si$^{+12}$  &   $5.0_{-0.5}^{+10}$  &   \nodata     &               &   \\
Si$^{+13}$  &   $8.0_{-1.6}^{+15}$    &   \nodata     &               &   \\

\enddata
\end{deluxetable}

\begin{deluxetable}{lccccc}
\tabletypesize{ \footnotesize }
 \tablecolumns{5} \tablewidth{0pt}
\tablecaption{Relative abundances with respect to Fe in the
ionized \x\ absorbers of \iras\ and \ngc, compared with solar
ratios \citep{as05} \label{tab2}} \tablehead{

   \colhead{Element} &
   %\colhead{Number}&
   \colhead{Solar Ratios} &
   \colhead{\iras} &
   \colhead{ \ngc} &
   \colhead{\iras} &
   \colhead{\ngc } \\
   \colhead{ }&
  % \colhead{\iras,\ngc }&
   \colhead{(A$_Z$/A$_{Fe}$)$_\odot$} &
   \colhead{(A$_Z$/A$_{Fe}$)} &
   \colhead{(A$_Z$/A$_{Fe}$) } &
   \colhead{(A$_Z$/A$_{Fe}$)/(A$_Z$/A$_{Fe}$)$_\odot$} &
   \colhead{(A$_Z$/A$_{Fe}$)/(A$_Z$/A$_{Fe}$)$_\odot$ }
}

\startdata

% Nitrogen& 2.1 $\pm$ 0.4  & $9.9_{-2.4}^{+81.2}$ &$6.4_{-0.6}^{+74.8} $ & $4.6_{-1.4}^{+38.0}$  & $3.0_{-0.6}^{+35.0} $ \\
% \cline{1-6}
O~~(5,5)\tablenotemark{a} & 16.2 $\pm$ 2.6  & $21.1_{-6.9}^{+6.8}$ & $16.5_{-2.3}^{+2.2} $ & $1.3_{-0.5}^{+0.5}$  & $1.0_{-0.2}^{+0.2} $ \\
\cline{1-6}
Ne (4,4) &  2.5 $\pm$ 0.4  & $3.3_{-1.0}^{+3.9}$ &$3.5_{-1.9}^{+0.8} $ & $1.3_{-0.4}^{+1.6} $ & $1.4_{-0.8}^{+0.3}$  \\
\cline{1-6}
% Sodium & 0.05 $\pm$ 0.01  & $0.7_{-0.2}^{+2.7} $&$0.10_{-0.04}^{+0.06} $ & $13.2_{-3.7}^{+51.0}$  & $1.9_{-0.8}^{+0.3}$  \\
% \cline{1-6}
Mg (8,9)&  1.2 $\pm$ 0.3  & $1.7_{-0.6}^{+0.6}$ &$1.9_{-0.7}^{+0.3}$  & $1.4_{-0.5}^{+0.5} $ & $1.5_{-0.7}^{+0.4}  $\\
\cline{1-6}
Si~~(9,9)& 1.1 $\pm$ 0.2  & $2.6_{-0.8}^{+1.0}$ &$2.1_{-0.6}^{+0.2} $ & $2.3_{-0.7}^{+0.8}$  & $1.9_{-0.6}^{+0.2}$  \\
\cline{1-6}
S~~~(0,2) & 0.5 $\pm$ 0.1  & \nodata &$0.5_{-0.2}^{+0.1}$  & \nodata  & $1.0_{-0.5}^{+0.2}$\\
%Iron& 1.0   & $0.96_{-0.12}^{+0.12} $&$1.01_{-0.05}^{+0.03} $  & $0.96_{-0.12}^{+0.12} $ & $1.01_{-0.05}^{+0.03} $  \\
%\cline{1-6}
\footnotesize \tablenotetext{a}{Number in parenthesis indicate the
number of ions used for each element, respectively for \iras\ and
for \ngc\ in the determination of the abundances.}

\enddata

\end{deluxetable}

\clearpage

\clearpage

\clearpage

\clearpage

\clearpage

\clearpage

\clearpage

\clearpage

%\epsscale{.70}

\end{document}